\documentclass[reprint, superscriptaddress, secnumarabic, amssymb, nobibnotes, aps]{revtex4-1}

\usepackage{graphicx}
\usepackage{epstopdf}
\usepackage[T1]{fontenc}
\usepackage[latin9]{inputenc}
\usepackage{amsbsy}
\usepackage{gensymb}
\usepackage[T1]{fontenc}
\usepackage[latin9]{inputenc}
\usepackage{amsmath}
\usepackage{amssymb}
\usepackage{bbm}
\usepackage{braket}
\usepackage{xcolor}
\allowdisplaybreaks
\usepackage{graphicx}
\usepackage[colorlinks=true]{hyperref}  
\hypersetup{
    bookmarks=true,         
    unicode=false,          
    pdftoolbar=true,        
    pdfmenubar=true,        
    pdffitwindow=false,     
    pdfstartview={FitH},    
    pdftitle={titel of paper},    
    pdfauthor={a},     
    pdfsubject={},   
    pdfcreator={},   
    pdfproducer={}, 
    pdfkeywords={} {} {}, 
    pdfnewwindow=true,      
    colorlinks=true,       
    linkcolor=blue, 
    citecolor=blue,        
    filecolor=magenta,      
    urlcolor=blue           
} 
\usepackage[normalem]{ulem}

\begin{document}
\title{\textrm{ High-temperature observation of intralayer, interlayer, and Rydberg excitons in \\
bulk van der Waals alloy single crystals }}
\author{Pravrati Taank}
\affiliation{Department of Physics, Indian Institute of Science Education and Research Bhopal, Bhopal 462066, India}
\author{Asif Ali}
\affiliation{Department of Physics, Indian Institute of Science Education and Research Bhopal, Bhopal 462066, India}
\author{Aravind Raji}
\affiliation{Department of Physics, Indian Institute of Science Education and Research Bhopal, Bhopal 462066, India}
\author{Ajay K. Poonia}
\affiliation{Department of Physics, Indian Institute of Science Education and Research Bhopal, Bhopal 462066, India}
\author{Matthew C. Beard}
\affiliation{Chemistry \& Nanoscience Center, National Renewable Energy Laboratory; Golden, Colorado 80401, USA}
\author{Ravi Shankar Singh}
\email[]{rssingh@iiserb.ac.in }
\affiliation{Department of Physics, Indian Institute of Science Education and Research Bhopal, Bhopal 462066, India}
\author{K. V. Adarsh}
\email[]{adarsh@iiserb.ac.in }
\affiliation{Department of Physics, Indian Institute of Science Education and Research Bhopal, Bhopal 462066, India}

\begin{abstract}
\begin{flushleft}

\end{flushleft}
Transition metal dichalcogenides (TMDs) exhibit remarkable optical properties due to the diverse number of strongly bound excitons, which can be fine-tuned by alloying. Despite a flurry of research activity in characterizing these excitons, a comprehensive and profound understanding of their behavior with temperature is lacking. Here, we report the rich spectrum of excitonic features within bulk van der Waals alloy Mo$_{0.5}$W$_{0.5}$S$_2$ and Mo$_{0.5}$W$_{0.5}$Se$_2$ single crystals through temperature-dependent reflectance spectroscopy and first-principle calculations. We observed Rydberg excitons and interlayer excitons in both the single crystals. Notably, we provide the first experimental evidence of highly energetic A$^\prime$ and B$^\prime$ excitons in Mo$_{0.5}$W$_{0.5}$S$_2$ at room temperature. The strong carrier-phonon scattering significantly broadens the A$^\prime$, B$^\prime$ and interlayer excitons at room temperature in bulk Mo$_{0.5}$W$_{0.5}$S$_2$ single crystal compared to its selenide. Our findings, supported by density functional theory and Bethe-Salpeter equation calculations, signify the crucial role of carrier-phonon interactions. These results open pathways for next-generation optoelectronic devices and quantum technologies operating at high temperature. 

\end{abstract}

\maketitle

Transition metal dichalcogenides (TMDs) have emerged as a new class of semiconductors for a wide range of applications in electronics and optoelectronics because of their remarkable properties like sizable bandgaps \cite{1}, substantial spin-valley coupling \cite{2,3}, spin-layer locking \cite{4,5}, reduced dielectric screening \cite{1,6}, and strong Coulomb interactions resulting in the high binding energy of excitons \cite{2,6,7}. These characteristics contribute to TMD's rich and diverse landscape of excitonic transitions, positioning them at the forefront of optical technology, particularly in the domains of spintronics \cite{8} and valleytronics \cite{9}. For example, large spin-orbit coupling (SOC) in TMDs splits the valance and conduction bands at the K-valley. This splitting results in the formation of low-energy intralayer A and B excitons and their excitonic Rydberg series \cite{2,10,11,12,13}. Another example is interlayer excitons (IXs) at the K-valleys, where electrons and holes residing in separate layers bind to form excitons based on spin-valley and spin-layer selection rules \cite{3,4,5,14}. Also, some of the TMDs, such as, selenides \cite{15,16,17,18} and tellurides \cite{15,18}, show highly energetic A$^\prime$ and B$^\prime$ excitons, however, these excitons are not observed in sulfides. For more details see Table S1 in Supplemental Material (SM) \cite{19} and in Refs. \cite{3,4,15,16,17,18,20,21,22,23,24,25,26}. Overall, all these diverse excitons present in TMDs offer a wide range of potential optoelectronic applications, including optical quantum sensing \cite{27}, Bose-Einstein condensation \cite{5,28}, optical switching \cite{12}, and quantum information technology \cite{29,30}. 
\par
TMDs present unique opportunities for alloying through isoelectronic substitution of transition metal (M) or chalcogen (X) atoms, resulting in minimal lattice strain and the formation of quasi-binary alloys that preserve the MX$_2$ structure \cite{31,32,33,34}. This structural stability allows diverse excitons within these quasi-binary alloys, mirroring those found in their binary counterparts. In principle, they can provide more fundamental insights than binary TMDs. For example, the right choice of atoms in an alloy offers a platform to tune the SOC and carrier-phonon scattering to explore a broader spectrum of excitons. Likewise, how the diverse excitons behave at different temperatures in TMDs is still lacking, which is essential for potential device applications.

Here, we present a comprehensive understanding of diverse excitonic features observed in the optical reflectance spectra of bulk Mo$_{0.5}$W$_{0.5}$S$_2$ and Mo$_{0.5}$W$_{0.5}$Se$_2$ alloy single crystals using temperature-dependent reflectance spectroscopy and first-principle calculations. In our experiments, we observed the Rydberg excitons of A exciton (1\textit{s} ground state) up to n=4 quantum number i.e., 2\textit{s}, 3\textit{s} and 4\textit{s} in Mo$_{0.5}$W$_{0.5}$S$_2$ single crystal. Another noteworthy findings are the first experimental observation of A$^\prime$ and B$^\prime$ excitons in transition metal sulfides at room temperature, and the observation of IX between A and B excitons in both single crystals. The strong carrier-phonon scattering significantly broadens these excitons at room temperature in bulk Mo$_{0.5}$W$_{0.5}$S$_2$ single crystal compared to its selenide. Furthermore, our theoretical calculations concur well with the experimental results. 
\par
Bulk Mo$_{0.5}$W$_{0.5}$S$_2$ and Mo$_{0.5}$W$_{0.5}$Se$_2$ alloy single crystals used in this study were procured from HQ Graphene. These crystals underwent extensive characterization (X-ray diffraction, scanning electron microscopy, energy-dispersive X-ray spectra and X-ray photoemission spectroscopy) to assess their structural and compositional properties. Details are provided in SM \cite{19}. Our measurements reveal the high quality and chemical purity of the single crystals.
\par
\begin{figure}[t]
\includegraphics[width=1.0\linewidth]{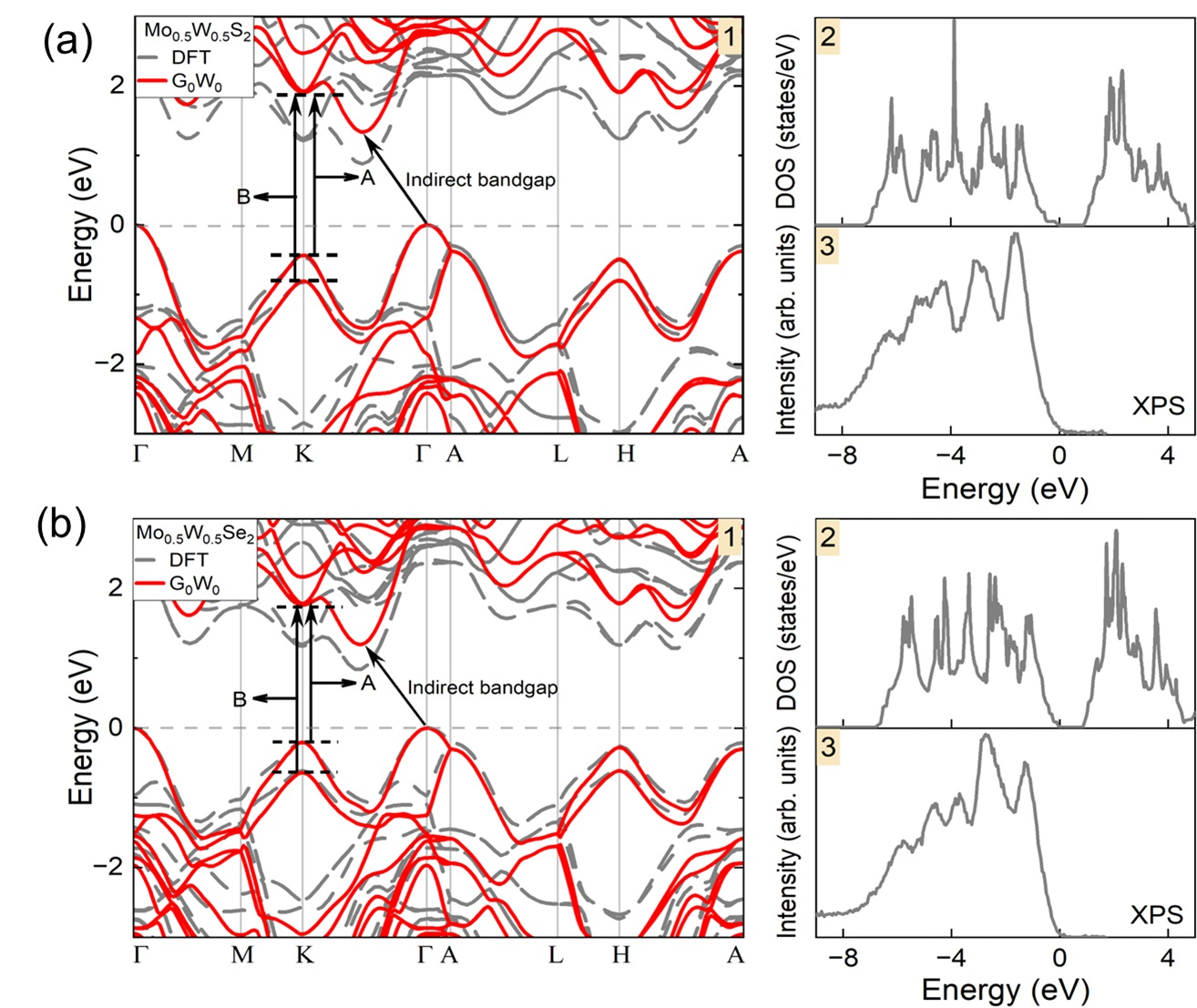} 
\caption{\label{fig:wide} 
Electronic band structure using DFT and G$_0$W$_0$ calculations of bulk (a) Mo$_{0.5}$W$_{0.5}$S$_2$ and (b) Mo$_{0.5}$W$_{0.5}$Se$_2$, (panel 1). The grey dashed line illustrates the Fermi levels. Panel 2 shows the calculated total DOS and panel 3 shows the room temperature experimental valence band spectra from XPS for both single crystals.}
\end{figure}
   Since the crystalline TMD of MX$_2$ displays large exciton binding energy and oscillator strength, we expect similar features in the optical and charge-transport properties of the alloy single crystals. For the identification of such diverse exciton species, we conducted a thorough investigation of the electronic structures of bulk Mo$_{0.5}$W$_{0.5}$S$_2$ and Mo$_{0.5}$W$_{0.5}$Se$_2$ within density functional theory (DFT) using the projector-augmented wave method \cite{35}  as implemented in the Vienna Ab-Initio Simulation Package (VASP) \cite{36} with the exchange-correlation functional of Perdew-Burke-Ernzerhof (PBE) \cite{37}. Experimental lattice parameters were used for all calculations, and virtual crystal approximation (VCA) \cite{38} was employed to represent the alloy compositions. The total density of state (DOS) calculations were done using the tetrahedron method. More details are in SM \cite{19} and in Refs. \cite{39,40,41,42}. The calculated electronic band structures and DOS of Mo$_{0.5}$W$_{0.5}$S$_2$ and Mo$_{0.5}$W$_{0.5}$Se$_2$ are shown in Figs. 1(a, b) in panels 1 and 2. Our calculations reveal that both alloys exhibit an indirect bandgap nature, with the respective valence band maximum (VBM) situated at the $\Gamma$ point and the conduction band minimum located between the K-$\Gamma$ points in both DFT (grey) and G$_0$W$_0$ (red) calculations. Additionally, the introduction of SOC leads to a splitting of the VBM, resulting in the first direct bandgap at the K point and the second direct bandgap at the H point. Our theoretical findings are consistent with the previously reported calculations in MoS$_2$ \cite{43} and WS$_2$ \cite{44} and highlight the significant contribution of transition metal d-electrons to the band structure of these bulk alloy single crystals.
\par
\begin{figure*}[htb]
\includegraphics[width=0.9\linewidth]{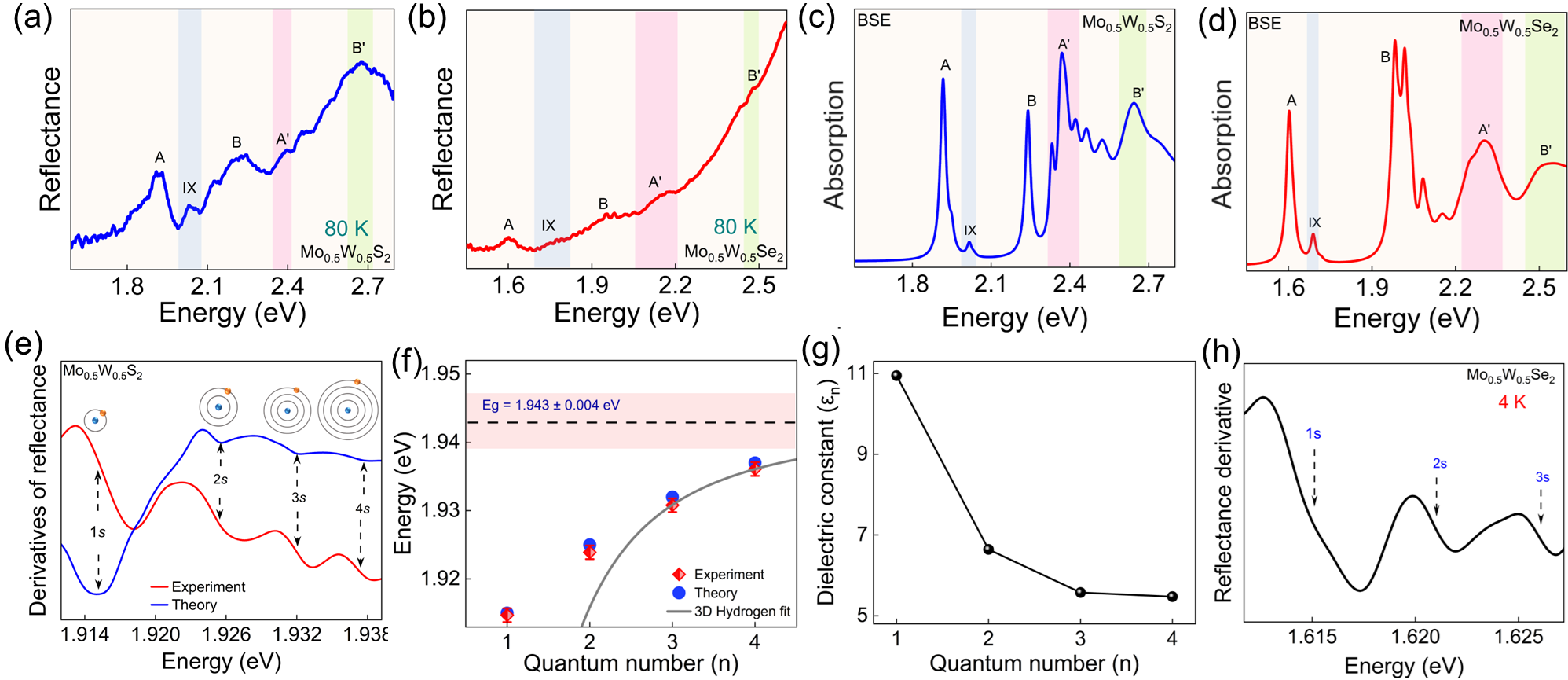}
\caption{\label{fig: wide} Reflectance spectrum at 80 K for bulk (a) Mo$_{0.5}$W$_{0.5}$S$_2$ and (b) Mo$_{0.5}$W$_{0.5}$Se$_2$ single crystals. The blue, pink, and green regions show the interlayer (IX), A$^\prime$ and B$^\prime$ excitons, respectively. The theoretical absorption spectrum calculated using BSE for bulk (c) Mo$_{0.5}$W$_{0.5}$S$_2$ and (d) Mo$_{0.5}$W$_{0.5}$Se$_2$. (e) The first derivative of the experimentally measured reflectance spectrum and the second derivative of the theoretically calculated BSE absorption spectrum of bulk Mo$_{0.5}$W$_{0.5}$S$_2$ single crystal at 80 K. The peaks represent the Rydberg excitons up to 4\textit{s} states, which follow Bohr's atomic model as shown schematically. Where blue and yellow balls are the nucleus (hole here) and the electron, respectively. The theoretical absorption spectra have been shifted by $\sim0.26$ eV for better comparison with the experimental spectra. (f) Experimentally and theoretically obtained the transition energies of the Rydberg series as a function of quantum number (n). The n=3,4 peaks follow the 3D hydrogen model and the fit is represented by the grey line. The black dashed line and pink shaded area show the estimated E$_g$ and uncertainty in the E$_g$ value, respectively. (g) Calculated dielectric constants $\epsilon_n$ as a function of n. (h) The second derivative of the reflectance spectrum of bulk Mo$_{0.5}$W$_{0.5}$Se$_2$ single crystal at 4 K shows Rydberg series up to n=3.}
\end{figure*}
\par
To understand the electronic structure of these systems, we compared the calculated DOS (panels 2) with the experimental valence band (panels 3) for bulk Mo$_{0.5}$W$_{0.5}$S$_2$ and Mo$_{0.5}$W$_{0.5}$Se$_2$ in Figs. 1(a, b), respectively. The observed valence band spectrum at room temperature exhibits semiconducting behavior with zero intensity at the Fermi level. The experimental valence band spectrum reveals five distinct features for Mo$_{0.5}$W$_{0.5}$S$_2$ and Mo$_{0.5}$W$_{0.5}$Se$_2$. These features reproduced in the calculated DOS are in good agreement with experimental valence band spectra. Furthermore, we compare the band dispersion with room temperature angle-resolved photoemission spectroscopy (ARPES), revealing the SOC-driven K-valley splitting of $\sim 0.37$ eV in Mo$_{0.5}$W$_{0.5}$Se$_2$, as shown in Fig. S6 \cite{19}. The DFT calculated band structure overlayed over the ARPES spectrum (Fig. S6 \cite{19}) shows a good agreement suggesting that DFT with VCA is sufficient enough to capture the main features of the valence band in the case of bulk alloy single crystals.
\par
   Bulk Mo$_{0.5}$W$_{0.5}$S$_2$ and Mo$_{0.5}$W$_{0.5}$Se$_2$ exhibit atomic stacking of $YY^*$ type, leading to 2H symmetry, as illustrated in Fig. S7 \cite{19}. These bulk single crystals demonstrate a wide range of bright intralayer and interlayer excitons with notably high oscillator strengths. The former obeys the spin-valley selection rules, while the latter follows the spin-layer selection rules \cite{4,45}. We conducted temperature-dependent reflectance spectroscopy to investigate these diverse optically active excitons and their impact on the optical properties of these single crystals. Detailed experimental procedure is in SM \cite{19}. Figs. 2(a, b) display the reflectance spectra at a high temperature of 80 K for bulk Mo$_{0.5}$W$_{0.5}$S$_2$ and Mo$_{0.5}$W$_{0.5}$Se$_2$ single crystals, respectively. These spectra clearly demonstrate the presence of multiple optically active excitons.  
\par
To gain a deeper understanding, we conducted theoretical calculations of the absorption spectra by incorporating electron-hole interactions and solving the Bethe-Salpeter equation (BSE) with the Tamm-Dancoff approximation \cite{46}. We considered the eight lowest conduction and eight highest valence bands for the BSE calculations. In Figs. 2(c, d), we present the theoretically calculated absorption spectra of bulk Mo$_{0.5}$W$_{0.5}$S$_2$ and Mo$_{0.5}$W$_{0.5}$Se$_2$, and the spectra are well in agreement with the experimental results in Figs. 2(a, b). These diverse spectral features observed in the experimental and theoretical spectra can be attributed to a range of excitonic transitions within the Brillouin zone, and their assignments are discussed in the following paragraphs.
\par
   First, we discuss the low-energy intralayer excitons, A and B, which are observed at $\sim1.915$ eV and $\sim2.221$ eV for Mo$_{0.5}$W$_{0.5}$S$_2$ and at $\sim1.603$ eV and $\sim1.955$ eV for Mo$_{0.5}$W$_{0.5}$Se$_2$. The A and B excitons originate from the splitting of the VBM due to strong SOC at K point of the Brillouin zone \cite{29,47}. The energy difference between the A and B excitons, which is an indicator of the strength of SOC \cite{21}, is $\sim306$ and $\sim352$ meV in Mo$_{0.5}$W$_{0.5}$S$_2$ and Mo$_{0.5}$W$_{0.5}$Se$_2$, respectively. These values are in good agreement with our theoretical calculations, for the respective samples $\sim323$ and $\sim378$ meV. The fact that Mo$_{0.5}$W$_{0.5}$Se$_2$ exhibits a stronger SOC, as indicated by the larger energy difference between A and B excitons, compared to Mo$_{0.5}$W$_{0.5}$S$_2$ is due to the heavier chalcogen atom. This is further supported by the band structure calculations presented in Fig. 1. The A and B exciton positions in our BSE calculation are slightly overestimated than the experimental values due to overestimation of bandgap in G$_0$W$_0$ calculations. 
   \par
   To unravel the subtle characteristics of other excitonic transitions between A and B excitons, we have taken the first derivative of the reflectance spectrum of Mo$_{0.5}$W$_{0.5}$S$_2$ single crystal and plotted it in Fig. 2(e). On the higher-energy side of the A exciton known as the 1\textit{s} ground state, we have obtained three additional peaks in Fig. 2(e). The intriguing aspect here lies in the gradual and continuous decrease in both peak intensities and energy separations as we move to higher energies is analogous to the excitonic Rydberg series \cite{2}. So, we associate these fine structures as the excited states of A exciton up to n=4 quantum number, i.e., 2\textit{s}, 3\textit{s}, and 4\textit{s}, since the energy separation and spectral weights resemble that of a hydrogen atom. Further, the transitions manifest as peaks situated between the 1\textit{s} ground state and the quasiparticle bandgap, as previously discussed in monolayer TMDs \cite{2,11,48}. These \textit{s}-like state transitions are dipole-allowed and thus can be probed by linear optical spectroscopy \cite{12}. Notably, our theoretical calculations also corroborate this trend. The second derivative of the theoretically calculated absorption spectrum exhibits a similar pattern, with the minima in the curve pinpointing the energy positions of the Rydberg excitons. We observed similar features using the first derivative of the reflectance data (Fig. S9 \cite{19}); however, the utilization of the second derivative enhances resolution, providing a more precise analysis. For clarity, we have outlined a schematic diagram in Fig. 2(e), aligning with Bohr's atomic model, where the electron resides in the corresponding shell. Furthermore, we have extracted the energy positions of these Rydberg excitons from the points of inflection of Fig. 2(e), and these positions are presented in Fig. 2(f) for a comprehensive view of the findings. Similar results are also obtained at 4 K (Fig. S10 \cite{19}); however, Rydberg excitons disappear as temperature increases beyond 80 K due to large carrier-phonon scattering \cite{49}. Further, we observed the Rydberg series of excitons in Mo$_{0.5}$W$_{0.5}$Se$_2$ at 4 K (Fig. 2(h)).
        \begin{figure*}[htb]
\includegraphics[width=0.8\linewidth]{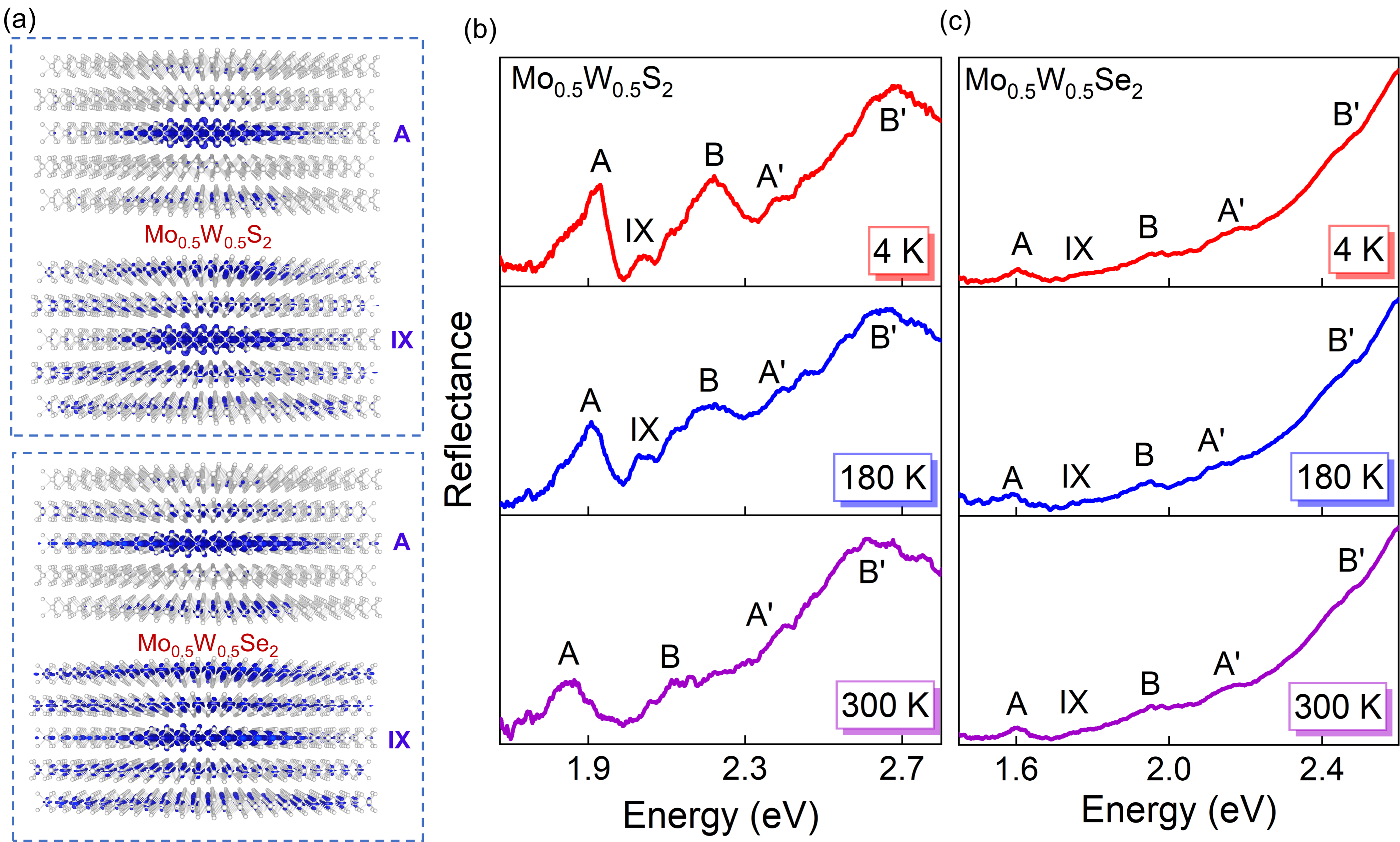}
\caption{\label{fig: wide}(a) Spatial distribution of excitons shows nature of A (intralayer) and IX excitons for Mo$_{0.5}$W$_{0.5}$S$_2$ (Top panel) and Mo$_{0.5}$W$_{0.5}$Se$_2$ (Bottom panel). Reflectance spectra as a function of temperature for bulk (b) Mo$_{0.5}$W$_{0.5}$S$_2$ and (c) Mo$_{0.5}$W$_{0.5}$Se$_2$ single crystals.}
\end{figure*}
\par
From the exciton Rydberg series of the single crystal, we have computed the exciton binding energy using the formula typically used for three-dimensional (3D) Wannier excitons \cite{50,51}, given by  $E_b^n= E_g-(R_y^*)/n^2$. Here, $E_b^n$, $E_g$, and $R_y^*$ are the binding energy of $n^{th}$ exciton, quasiparticle bandgap, and effective Rydberg constant, respectively. We have used only n=3 and 4 Rydberg excitons for fitting the data, since they are hydrogenic in nature, which is discussed later. The extracted value of $E_g$ = 1.943 $\pm$ 0.004 eV (black dashed line in Fig. 2(f)) and $E_b (= E_g-E_{A(1s)})  = 28$
 $\pm$ 4 meV. Notably, our observed binding energy values are similar to those of GaSe \cite{52}, GaN \cite{53} and ZnO \cite{54} bulk single crystals, but at least an order of magnitude smaller than monolayer MoS$_2$ and WS$_2$ \cite{2,10} due to the significantly large screening effect in the bulk material. To explain the observed non-hydrogenic behavior, we determined the non-uniform dielectric constant $\epsilon_n=\sqrt{\mu e^4\ /(2\hbar^2 E_b^{(n)} n^2)}$ experienced by the Rydberg excitons \cite{2}, where $\mu$ represents the exciton reduced mass in terms of the electron rest mass (m$_0$) and is found to be 0.2471m$_0$, as estimated by DFT calculations at the K-valley. Our results demonstrate a drastic reduction in $\epsilon_n$ for n = 1 and 2, while it remains nearly constant for n = 3, 4 Rydberg excitons as shown in Fig. 2(g). This justifies the use of the 3D hydrogenic model for fitting n = 3, 4, and the deviation from the hydrogenic model for n =1, 2 can be attributed to the antiscreening effect, consistent with previous reports in other TMDs \cite{2,10} and semiconductors \cite{55,56}. 
 \par
   Another remarkable observation of our study is the identification of a distinctive spectral feature appearing at high energy to the A exciton by $\sim114$ meV in Mo$_{0.5}$W$_{0.5}$S$_2$ and $\sim156$ meV in Mo$_{0.5}$W$_{0.5}$Se$_2$, which is well below the B exciton within the reflectance spectrum, see Figs. 2(a, b). We assign these features to the direct IXs. Notably, our temperature-dependent studies presented in Fig. 3(b) show the weakening of IX peak after 180 K in Mo$_{0.5}$W$_{0.5}$S$_2$, nevertheless, in Mo$_{0.5}$W$_{0.5}$Se$_2$, we could observe IX up to room temperature (Fig. 3(c)). In contrast to the direct intralayer A exciton, these IXs exhibit reduced oscillator strength, strong dipole-dipole interactions, and lower binding energy. To provide a clearer understanding of the underlying mechanism, we have shown the schematic representation of band structure (Fig. S11 \cite{19}) and spin configurations of IXs at the K-valley in the $YY^*$ stacked bulk alloy single crystal, considering two adjacent layers as depicted in Fig. S12 \cite{19}. At each K point, the band structure is composed of SOC split valence bands $V_1^{(\uparrow)}$ and $V_2^{(\downarrow)}$, as well as conduction bands $C_1^{(\uparrow)}$, and $C_2^{(\downarrow)}$ in layer 1 and with spin inversion in layer 2, referred to as spin-layer locking \cite{4,5}. Due to spin conservation in optical dipole transitions, IXs form within the same spin-oriented bands of adjacent layers \cite{4}. To confirm that the assigned peaks are IXs, we have calculated the exciton wave function which shows the charge density plot of the electron by keeping the hole position fixed very close to the transition metal position, as shown in Fig. 3(a). As can be noticed from the figure that the A exciton is confined within one layer, suggesting intralayer exciton like nature. On the other hand, IX is delocalized in neighboring layers, indicating interlayer characteristics. 
\par
Likewise, we observe IX in Mo$_{0.5}$W$_{0.5}$Se$_2$ single crystal, see Fig. 3(c); however, a relatively broad and weak feature of IX is detected. This can be attributed to the weak interlayer coupling between the bands due to the larger spin-orbit splitting of the electronic bands \cite{3,57}, i.e., Mo$_{0.5}$W$_{0.5}$Se$_2$ exhibits larger spin-orbit splitting than Mo$_{0.5}$W$_{0.5}$S$_2$, see Figs. 3(b, c). This fosters stronger interlayer coupling, forming IX of larger oscillator strength in Mo$_{0.5}$W$_{0.5}$S$_2$ compared to Mo$_{0.5}$W$_{0.5}$Se$_2$. In contrast to Mo$_{0.5}$W$_{0.5}$Se$_2$, this IX becomes less discernible at room temperature in Mo$_{0.5}$W$_{0.5}$S$_2$. This may be due to the delocalization of carriers caused by high carrier-phonon scattering in sulfides, as noticed from high LO phonon energy in our phonon calculations discussed later. Further, we extended our analysis to include monolayers for the completeness of our theoretical calculation. As expected, in the case of monolayers, we did not observe any signature of the IX, see Fig. S13 \cite{19}.
\par
Another noteworthy observation of our study is the demonstration of highly energetic A$^\prime$ and B$^\prime$ excitons originating from SOC split valance band at K-valley (V$_1$ and V$_2$) to the higher conduction band since the energy separation of these features are same as that of A and B excitons (see Fig. S11 \cite{19}). While the experimental identifications of these A$^\prime$ and B$^\prime$ excitons have been previously reported in materials like MoSe$_2$ \cite{18}, WSe$_2$ \cite{16,17}, and MoTe$_2$ \cite{15,18}, it is important to note that their presence has not yet been experimentally demonstrated in any sulfide TMDs in monolayer or bulk, despite the theoretical predictions \cite{7,58}. The theoretical analysis suggests that the large size of anion atoms is not only responsible for the observation of A$^\prime$ and B$^\prime$ excitons but also for influencing the electron-phonon interactions depending on the excitonic energies \cite{7}. This means at higher temperatures, electron-phonon interactions strongly affect the higher-energy excitons than the lower-energy excitons. As a result, higher energy excitonic peaks merge into a much broader peak and so could not be resolved in the experiments at relatively high temperatures. We conducted comprehensive temperature-dependent reflectance measurements from 4 to 300 K to address this, as presented in Figs. 3(b, c). This analysis revealed that at the lowest temperature of 4 K, we could distinctly observe the sharp A$^\prime$ and B$^\prime$ excitons in bulk Mo$_{0.5}$W$_{0.5}$S$_2$ single crystal. However, as the temperature increased, they coalesced into much broader peaks and become less prominent at room temperature. Contrastingly, A$^\prime$ and B$^\prime$ excitons are observed in bulk Mo$_{0.5}$W$_{0.5}$Se$_2$ single crystal are less affected with temperature, Fig. 3(c). To understand the mechanism and role played by phonons in the broadening of A$^\prime$ and B$^\prime$ excitons, we have calculated the phonon dispersion curves of both alloys in Fig. S14 \cite{19}. More details are in SM  \cite{19} and Refs.  \cite{59,60,61}. Due to the heavier atomic weight of Se atoms, Mo$_{0.5}$W$_{0.5}$Se$_2$ phonon bands are shifted downwards to the lower frequencies compared to their sulfides. This signifies that carrier-phonon interactions become more significant in Mo$_{0.5}$W$_{0.5}$S$_2$ due to high LO phonon energy of $\sim45$ meV in comparison to Mo$_{0.5}$W$_{0.5}$Se$_2$ ($\sim32$ meV). As a result, highly energetic A$^\prime$ and B$^\prime$ excitons in Mo$_{0.5}$W$_{0.5}$S$_2$ are becoming much broader and weaker at higher temperatures, but clearly observed as sharper peaks at low temperatures. In contrast, weaker carrier-phonon interaction in Mo$_{0.5}$W$_{0.5}$Se$_2$ is not affecting these excitons at room temperature similar to previous reports of transition metal selenides \cite{15,16,17,18} and tellurides \cite{15,18}. \par
In conclusion, temperature-dependent reflectance spectroscopy, in conjunction with theoretical calculations, demonstrates various optically active transitions in bulk Mo$_{0.5}$W$_{0.5}$S$_2$ and Mo$_{0.5}$W$_{0.5}$Se$_2$ alloy single crystals. Apart from intra-, inter-layer and highly energetic A$^\prime$ and B$^\prime$ excitons, we also observe Rydberg excitons in both systems. Temperature-dependent studies show that at higher temperatures electron-phonon interactions strongly affect the higher-energy excitons than the lower-energy excitons. Our comprehensive study reveals these exotic excitonic states on a single platform, which will help to understand the fundamental physics and may open up the opportunities for the implementation of Bose-Einstein condensation, quantum information processing and optoelectronic applications at room temperature.\par
\textbf{Acknowledgments$:$}
Authors gratefully acknowledge the Department of Science \& Technology (Project no: DST/NM/TUE/QM-8/2019 (G)/1) and Science and Engineering Research Board (Project no: CRG/2019/002808) for financial support. We thankfully acknowledge central instrumentation and HPC facilities at IISER Bhopal.

\begin{widetext}
	\includegraphics[page=1,width=\textwidth]{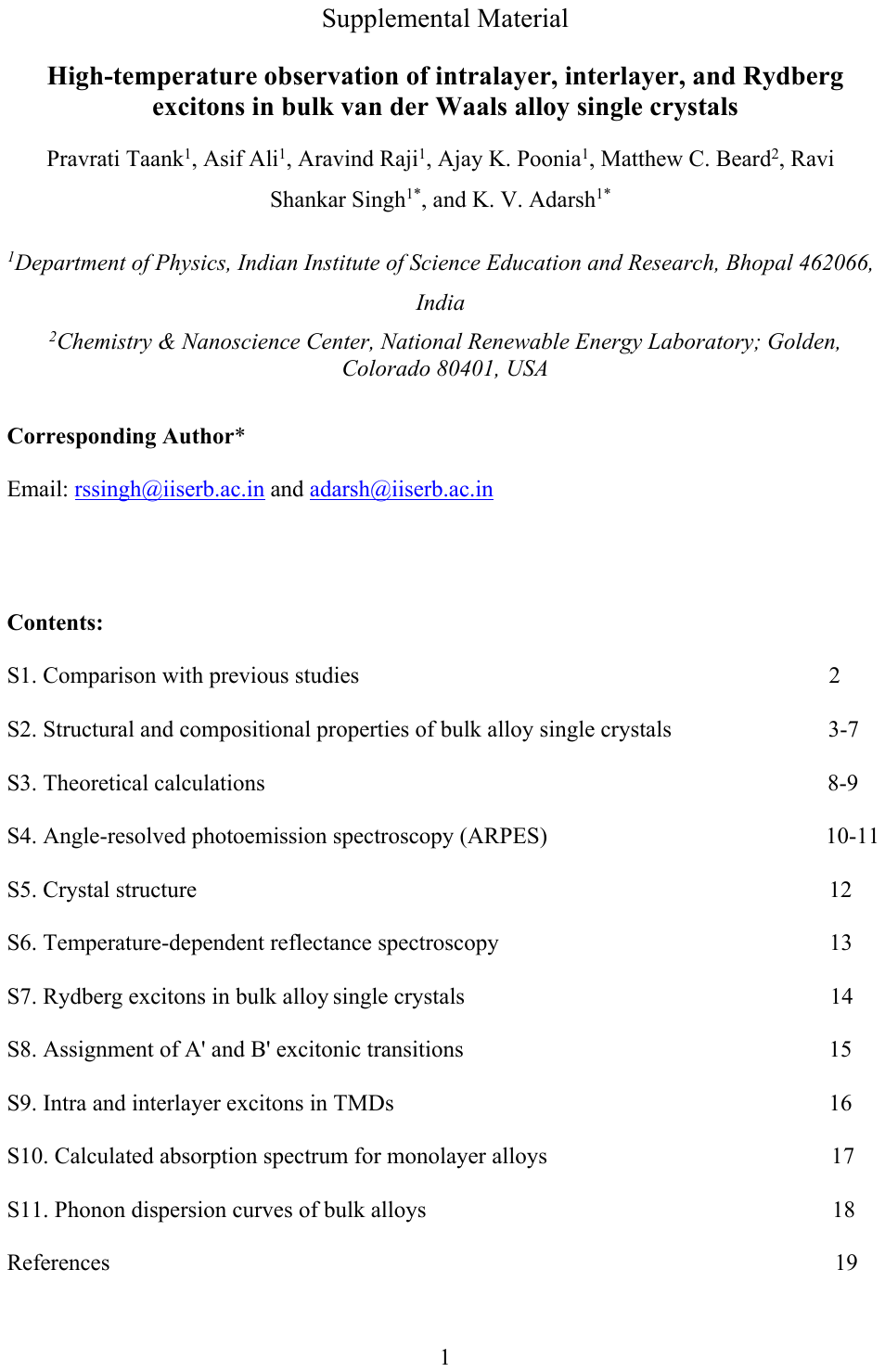}
	\includegraphics[page=2,width=\textwidth]{SM.pdf}
	\includegraphics[page=3,width=\textwidth]{SM.pdf}
	\includegraphics[page=4,width=\textwidth]{SM.pdf}
	\includegraphics[page=5,width=\textwidth]{SM.pdf}
	\includegraphics[page=6,width=\textwidth]{SM.pdf}
	\includegraphics[page=7,width=\textwidth]{SM.pdf}
	\includegraphics[page=8,width=\textwidth]{SM.pdf}
	\includegraphics[page=9,width=\textwidth]{SM.pdf}
	\includegraphics[page=10,width=\textwidth]{SM.pdf}
	\includegraphics[page=11,width=\textwidth]{SM.pdf}
	\includegraphics[page=12,width=\textwidth]{SM.pdf}
	\includegraphics[page=13,width=\textwidth]{SM.pdf}
	\includegraphics[page=14,width=\textwidth]{SM.pdf}
	\includegraphics[page=15,width=\textwidth]{SM.pdf}
	\includegraphics[page=16,width=\textwidth]{SM.pdf}
	\includegraphics[page=17,width=\textwidth]{SM.pdf}
	\includegraphics[page=18,width=\textwidth]{SM.pdf}
	\includegraphics[page=19,width=\textwidth]{SM.pdf}
\end{widetext}

\end{document}